\documentclass[numberedappendix]{emulateapj}
\usepackage{graphicx,amssymb,amsmath,epsfig}
\renewcommand{\aj}{AJ}\renewcommand{\apj}{ApJ}\renewcommand{\apjl}{ApJL}\renewcommand{\apjs}{ApJS}
\renewcommand{\mnras}{MNRAS}\renewcommand{\nat}{Nature}\renewcommand{\prd}{PhRvD}\renewcommand{\prl}{PhRvL}\newcommand{\ijmpd}{IJMPD}
\newcommand{\msol}{\mbox{M$_{\odot}$}}\newcommand{\rs}{$r_{\rm{s}}$}\newcommand{\rc}{$r_{\rm{c}}$}\newcommand{\rhoo}{$\rho_{\rm{o}}$}
\newcommand{\rmatch}{$r_{\rm{m}}$}\newcommand{\rh}{$r_{\rm{h}}$}\newcommand{\Mh}{$M_{\rm{h}}$}\newcommand{\Ms}{$M_{\rm{s}}$}\newcommand{\Wh}{$W_{\rm{h}}$}\newcommand{\Wc}{$W_{\rm{c}}$}\newcommand{\DE}{$\Delta E$}
\begin{document}
\title{The Energetics of Cusp Destruction}
\shorttitle{The Energetics of Cusp Destruction}
\author{Aaron J. Maxwell, James Wadsley, \& H. M. P. Couchman}
\shortauthors{Maxwell, Wadsley, \& Couchman}
\affil{Department of Physics \& Astronomy, McMaster University, Hamilton, ON, L8S 4M1, CAN}\email{ajmax@mcmaster.ca}
\begin{abstract}
We present a new analytic estimate for the energy required to create a constant density core within a dark matter halo.
Our new estimate, based on more realistic assumptions, leads to a required energy that is orders of magnitude lower than is claimed in earlier work.
We define a core size based on the logarithmic slope of the dark matter density profile so that it is insensitive to the functional form used to fit observed data.
The energy required to form a core depends sensitively on the radial scale over which dark matter within the cusp is redistributed within the halo.
Simulations indicate that within a region of comparable size to the active star forming regions of the central galaxy that inhabits the halo, dark matter particles have their orbits radially increased by a factor of 2--3 during core formation.
Thus the inner properties of the dark matter halo, such as halo concentration, and final core size, set the energy requirements.
As a result, the energy cost increases slowly with halo mass as \Mh$^{0.3-0.7}$ for core sizes $\lesssim$\thinspace1\thinspace kpc.
We use the expected star formation history for a given dark matter halo mass to predict dwarf galaxy core sizes.
We find that supernovae alone would create well over 4\thinspace kpc cores in $10^{10}$\thinspace\msol~dwarf galaxies \emph{if} 100\% of the energy were transferred to dark matter particle orbits.
We can directly constrain the efficiency factor by studying galaxies with known stellar content and core size, such as Fornax.
We find that the efficiency of coupling between stellar feedback and dark matter orbital energy need only be at the 1\% level or less to explain Fornax's 1 kpc core.
\end{abstract}
\section{Introduction}
\indent Collisionless simulations of dark matter halos in a $\Lambda$CDM cosmology consistently predict a density profile that diverges towards the centre \citep[e.g.][]{dubinski1991, navarro1995, navarro1997, bullock2001, klypin2001, stadel2009} over a wide mass range.
This profile typically rolls over to a steep slope at large radii where it is broadly consistent with the rotation curves of many late-type galaxies \citep[e.g.][]{rubin1980, bosma1981}.
However, the inner cusp \citep[e.g. $\rho\propto r^{-1}$,][]{navarro1995} is \emph{inconsistent} with observations of rotation profiles in dwarf galaxies both within the local group and beyond \citep[e.g.][]{flores1994, moore1994, burkert1995, cote2000, kuziodenaray2006, gilmore2007, kuziodenaray2008, oh2011a, walker2011, amorisco2013, adams2014}.
Instead, these dwarf galaxies are better fit using a constant density within the central kpc or so.
Two classes of solutions have been proposed to resolve this discrepancy.
The first suggests that coupling to baryons can reshape the dark matter cusp into a core \citep[e.g.][]{navarro1996b, elzant2001, weinberg2002, read2005, mashchenko2006}.
These solutions have met with some criticism \citep[e.g.][]{sellwood2003, tasitsiomi2003, jardel2009, dubinski2009}.
In particular, \citet{penarrubia2012} claimed it was not energetically feasible.
However, the sub-structure present within massive halos can induce collisionless scattering of dark matter particles within the centre, flattening the inner cusp without the need of baryons \citep[e.g.][]{ma2004}.
Alternative models to $\Lambda$CDM have been also been suggested, such as `warm' dark matter \citep[e.g.][]{hogan2000} or (non-gravitational) dark matter self-interactions \citep[e.g.][]{spergel2000}.\\
\indent \citet{mashchenko2006} argued that cyclic star formation bursts within the centres of dark matter halos could provide the necessary energy to re-shape the dark matter cusp into a dark matter core.
In this model, feedback from type II supernovae (SNe) periodically drives bulk gas motions in the star forming regions in the central kpc of dwarf galaxies.
Since gas can dominate the central gravitational field, these bulk motions can significantly perturb the central gravitational potential on timescales comparable to dark matter orbital times leading to an efficient transfer of kinetic energy to the central dark matter distribution and the formation of a significant dark matter core.
A notable feature of the \citet{mashchenko2006} framework was that the gas need not be expelled or even pushed very far outside the core-forming region \citep[as opposed to the impulsive blow-out scheme e.g.][]{navarro1996b, read2005}.
\citet{pontzen2012} gave an analytical description of how changing the dominate potential leads to irreversible heating of the dark matter orbits.\\
\indent The mechanism was first verified in a cosmological simulation by \citet{mashchenko2008}.
A required feature for simulations to flatten cusps is realistically clustered star formation so as to ensure a star formation history that significantly fluctuates in both space and time.
Several simulations since have reproduced these results over a range of simulated galaxy masses \citep[e.g.][see \citeauthor{pontzen2014} \citeyear{pontzen2014}]{governato2010, governato2012, zolotov2012, teyssier2013, madau2014}.
The process also has consequences for other collisionless components such as stars and star clusters.
Initially centrally concentrated populations (such as younger stars) are pushed outward over time, leading to older populations having larger orbits on average.
This results in a broad match to observations of the stellar content of dwarf galaxies \citep[e.g.][]{maxwell2012, teyssier2013, governato2014}.
It also has interesting consequences for multiple populations in galactic globular clusters \citep[][]{maxwell2014}.\\
\indent \citet{penarrubia2012} made the first analytical estimate of the energy difference between cored and cuspy dark matter halos of the same mass, finding that SNe alone would have difficulty providing the energy required to form dark matter cores given the average stellar masses of dwarf galaxies.
Though there are other sources of energy available from star formation, such as stellar winds and the ionizing flux from young stars \citep[e.g.][]{leitherer1999}, it is not clear that this added energy would be sufficient to remove the discrepancy or couple as well as SNe-driven bulk gas motions.
However, the \citet{penarrubia2012} results are difficult to reconcile with the core forming simulations which typically employ feedback only from SNe (and with realistic efficiencies).\\
\indent In this work we re-examine the energy requirements and show that the formation of dark matter cores is energetically plausible in a $\Lambda$CDM cosmology.
The layout of the paper is as follows.
\S\ref{sec:amb} deals with the ambiguity inherent in the definition of a core size, and argues for a new, general way to define the core radius.
In \S\ref{sec:cfeb} we discuss the energy required to form dark matter cores in halos that were initially cuspy using assumptions that are more consistent with our understanding of how the process operates.
In \S\ref{sec:results} we show predictions for core sizes in dark matter halos and examine the implications for observed systems such as Fornax.
\section{Consistent Definition of Core Size}\label{sec:amb}
\begin{figure*}
	\centering
	\plottwo{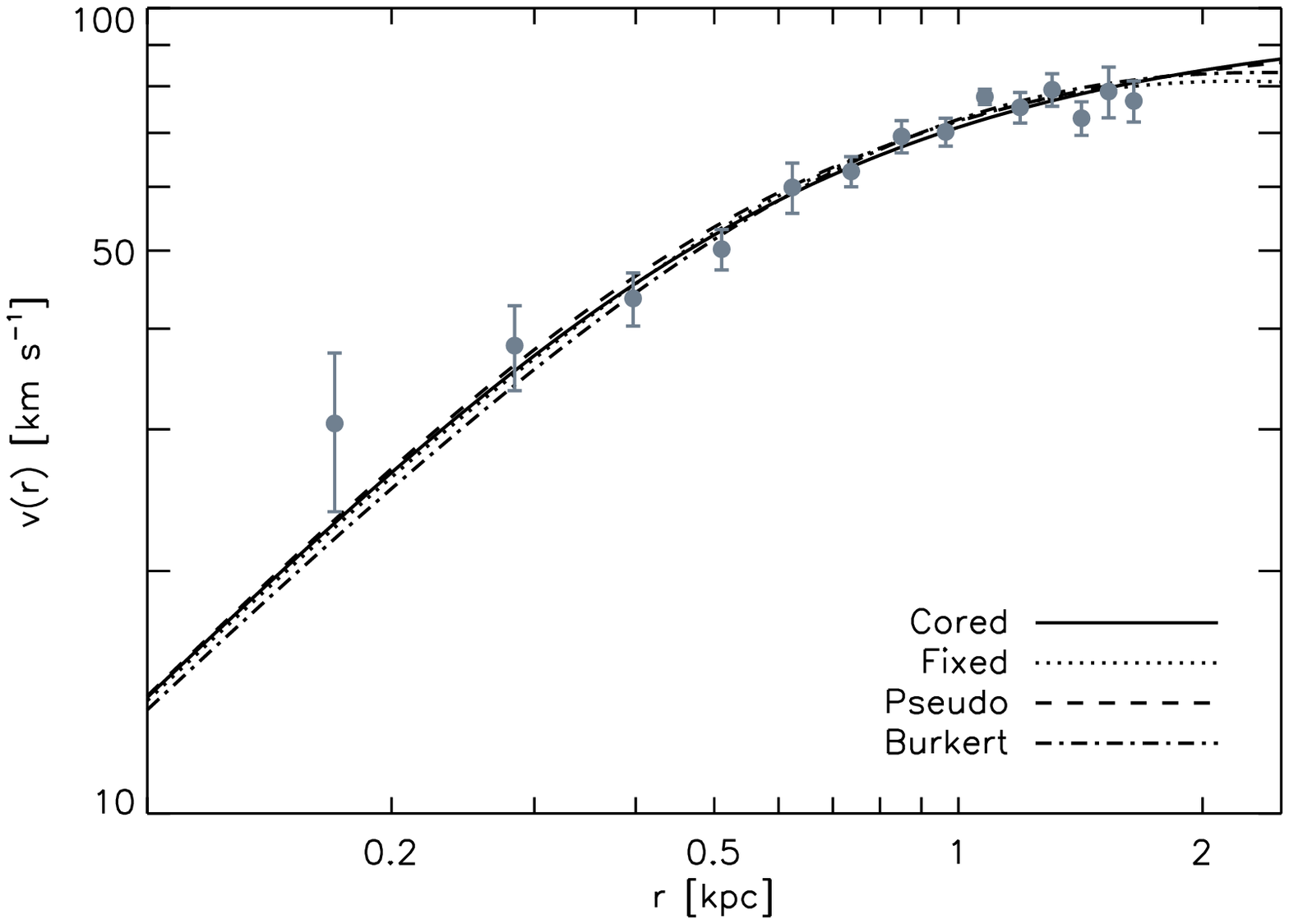}{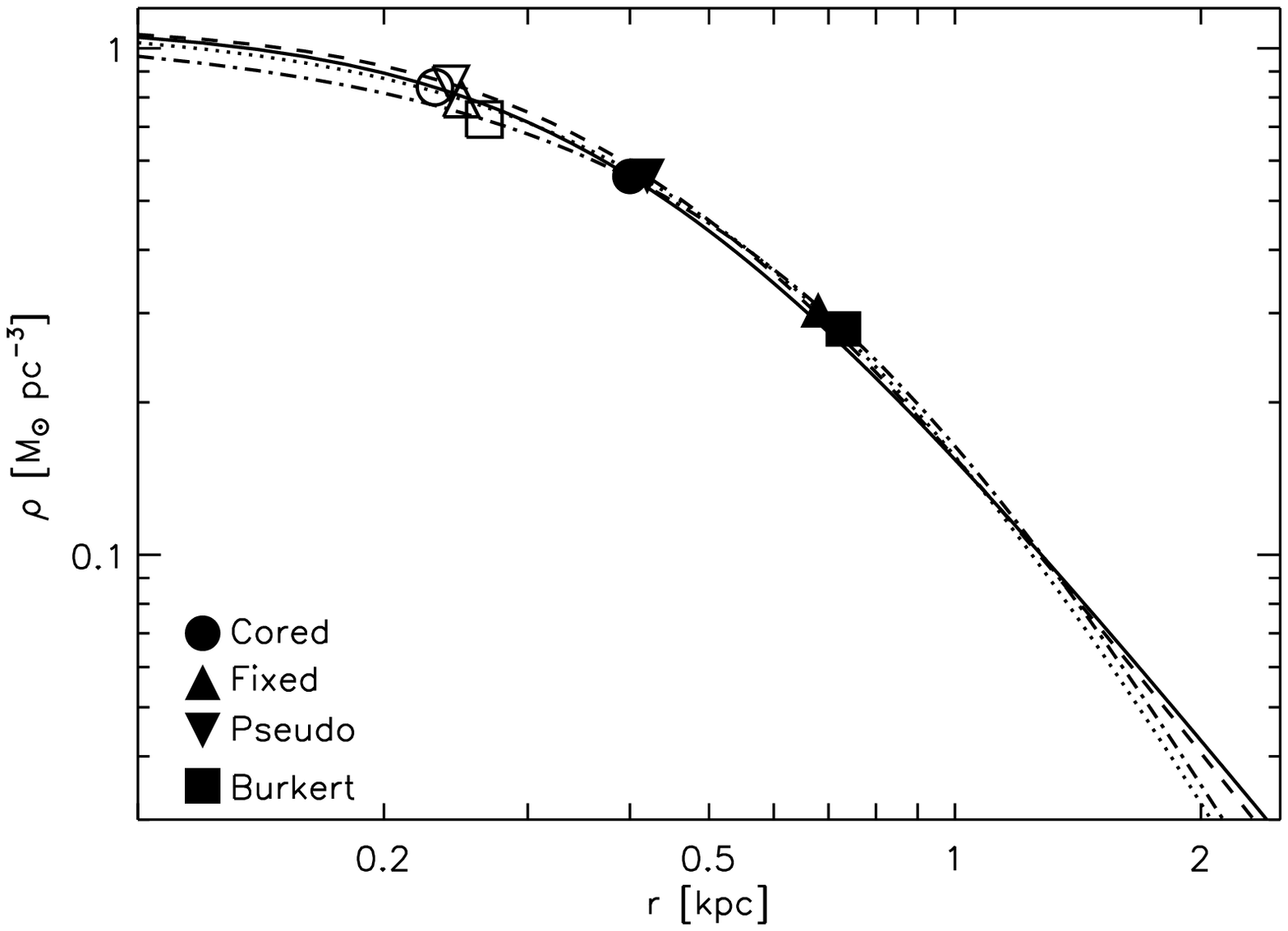}
	\caption[kdn]{\emph{LEFT}: The velocity profile of the LSB NGC 959 from \citet{kuziodenaray2008}, with the best fit `zero disk' cored isothermal profile shown as the solid line.  The dotted line shows the best fit using Equation \eqref{eq:rhofix}, the dashed line shows Equation \eqref{eq:rhocore} and the dash-dot line shows the best fit using the \citet{burkert1995} profile (Eqn. \ref{eq:rhoburk}).  For many of these halos, differences between the cored profiles are only evident at radii well outside the available observations.  \emph{RIGHT}: The dark matter density profiles derived from the best fit velocity profiles.   For the same observed dwarf galaxy four different scale radii, the traditional definition of the core size, can be derived.  The filled symbols show \rs~for each of the four profiles, while the open symbols show \rc, where $\rm{d}\ln{\rho}/\rm{d}\ln{r}=-1/2$.}
	\label{fig:kdn}
\end{figure*}

\indent Many observed cores in dwarf galaxies are inferred from the rotation profiles of the stars and gas by finding the best-fit dark-matter density model \citep[e.g.][]{flores1994, burkert1995, kuziodenaray2006, gilmore2007, kuziodenaray2008, oh2011a}.
Commonly, one would use either a cored isothermal sphere, whose density profile goes as $r^{-2}$ at large radii, or that proposed by \citet{burkert1995} where the density goes as $r^{-3}$ at large radii,
\begin{equation}\label{eq:rhoburk}
	\rho(r)\propto\frac{1}{\big(1+(\frac{r}{r_{\rm{s}}})\big)\big(1+(\frac{r}{r_{\rm{s}}})^{2}\big)}.
\end{equation}
Both satisfy the requirement that the density profile transitions to a flat core within a scale radius.
Without probing the rotation profile to large radii it is difficult to determine which density profile is the best fit.
The cored isothermal profile predicts a constant velocity at large radii while the \citet{burkert1995} profile looks like the standard galaxy rotation curve \citep[e.g.][]{rubin1978}.
However, the common practice of defining the core size as the scale radius leads to large variations in core radii dependent on the chosen functional form.\\
\indent To avoid this ambiguity, we define the core radius as \emph{the radius, \rc, at which the logarithmic slope of the density profile satisfies}:
\begin{equation*}
	\frac{\rm{d}\ln{\!\rho}}{\rm{d}\ln{\!r}}=-\frac{1}{2}.
\end{equation*}
We will use this definition from now on when referring to the core size.
Characterizing cores this way is consistent with the density profile slopes found in simulations of dwarf galaxies \citep[e.g.][]{mashchenko2008,governato2010,teyssier2013} and with the mass profile slopes observed in local group dwarfs \citep[e.g.][]{walker2011,amorisco2013,adams2014}.
Furthermore, the simulated galaxies always transition to a density profile with a strongly negative slope ($\rm{d}\ln{\rho}/\rm{d}\ln{r}\simeq-3$) beyond a few kpc where the effects of baryonic feedback are limited.
This definition of the core radius lessens the reliance on an assumed dark matter density distribution, and it can be determined directly from the velocity profile.
It is also straightforward to relate \rs~(for any given density profile) to \rc.\\
\indent To demonstrate the robustness of our new core definition, and to illustrate the ambiguity with using the scale radius, we show in Figure \ref{fig:kdn} the rotational velocity profile for the Low Surface Brightness (LSB) galaxy NGC 959 from \citet{kuziodenaray2008}.
These authors assumed a cored isothermal density profile, and the best fit of this profile to the data is shown as the solid line.
Although it is perfectly sufficient to match the profile in the inner 1--2\thinspace kpc, it does not roll over sufficiently to match the slope in the outer part of a cosmological halo.
We have also fit three other density profiles.
The first, shown as the dotted line in the figure, is \citep[e.g.]['Fixed' in the figure]{widrow2000}:
\begin{equation}\label{eq:rhofix}
	\rho(r)\propto\frac{1}{\big(1+(\frac{r}{r_{\rm{s}}})^{\alpha}\big)^{3/\alpha}}.
\end{equation}
The second, shown as the dashed line, is a pseudo-isothermal profile, where we let the power index vary:
\begin{equation}\label{eq:rhocore}
	\rho(r)\propto\frac{1}{1+(\frac{r}{r_{\rm{s}}})^{\delta}}.
\end{equation}
Lastly, we show the \citet[Eqn. \ref{eq:rhoburk}]{burkert1995} profile as the dash-dot line.\\
\indent The differences between the three rotation profile fits are only evident at radii past the furthest observable point, where the predicted rotation velocity rolls over.
We have also done this for the eight other LSBs from \citet{kuziodenaray2008} (see the Appendix), and the resulting scale radii and core radii are shown in Table \ref{tab:kdnm}.
It is evident that velocity rotation data alone are not sufficient to distinguish the best fit dark matter density profile.
Each fitting function requires a different scale radius, the common definition of the core size.
However, using $\rm{d}\ln{\!\rho}/\rm{d}\ln{\!r}$ to define the core radius leads to consistent core sizes, independent of which of the four density profiles was used, so long as the core location is comfortably within the range of data points.\\
\indent There are some LSBs in the \citet{kuziodenaray2008} sample that do not show consistent \rc~values, particularly UGC~4325 and DDO~64.
These are galaxies where the core size is beyond the edge of the data, so a fitting algorithm is only able to place lower bounds on \rc~and is certainly not able to constrain the parameters of a non-linear density function.
\begin{deluxetable*}{cccc|ccc|ccc|cc}
	\tabletypesize{\scriptsize}
	\tablecolumns{12}
	\tablewidth{0pt}
	\tablecaption{Best fit parameters.}
	\tablehead{\colhead{ }&\multicolumn{3}{c}{Cored}&\multicolumn{3}{c}{Pseudo}&\multicolumn{3}{c}{Fixed}&\multicolumn{2}{c}{Burkert}\\\colhead{Galaxy}&\colhead{$\delta$}&\colhead{\rs}&\colhead{\rc}&\colhead{$\delta$}&\colhead{\rs}&\colhead{\rc}&\colhead{$\alpha$}&\colhead{\rs}&\colhead{\rc}&\colhead{\rs}&\colhead{\rc}}
	\startdata
	{\small NGC 959}&{\small 2}&{\small 0.4}&{\small 0.23}&{\small 2.10}$_{-0.18}^{+0.18}$&{\small 0.42}$_{-0.02}^{+0.02}$&{\small 0.24}&{\small 1.60}$_{-0.17}^{+0.19}$&{\small 0.68}$_{-0.05}^{+0.07}$&{\small 0.25}&{\small 0.73}$_{-0.01}^{+0.01}$ &{\small 0.27}\\[0.3em]
	{\small NGC 7137}&{\small 2}&{\small 0.6}&{\small 0.35}&{\small 2.01}$_{-0.34}^{+0.38}$&{\small 0.61}$_{-0.10}^{+0.08}$&{\small 0.35}&{\small 1.41}$_{-0.36}^{+0.52}$&{\small 1.14}$_{-0.22}^{+0.44}$&{\small 0.36}&{\small 1.09}$_{-0.04}^{+0.04}$&{\small 0.40}\\[0.3em]
	{\small UGC 11820}&{\small 2}&{\small 1.1}&{\small 0.64}&{\small 3.84}$_{-1.14}^{+1.93}$&{\small 1.15}$_{-0.06}^{+0.06}$&{\small 0.70}&{\small 2.76}\tablenotemark{a}&{\small 0.89}$_{-0.04}^{+0.31}$&{\small 0.50}&{\small 2.11}$_{-0.23}^{+0.28}$&{\small 0.77}\\[0.3em]
	{\small UGC 128}&{\small 2}&{\small 2.3}&{\small 1.33}&{\small 1.97}$_{-0.32}^{+0.36}$&{\small 2.48}$_{-0.37}^{+0.32}$&{\small 1.43}&{\small 1.32}$_{-0.27}^{+0.37}$&{\small 5.01}$_{-0.94}^{+1.63}$&{\small 1.48}&{\small 4.50}$_{-0.16}^{+0.16}$&{\small 1.64}\\[0.3em]
	{\small UGC 191}&{\small 2}&{\small 1.7}&{\small 0.98}&{\small 1.64}$_{-0.64}^{+1.16}$&{\small 1.84}$_{-0.28}^{+0.85}$&{\small 1.11}&{\small 1.48}$_{-0.48}^{+1.32}$&{\small 3.37}$_{-1.67}^{+6.26}$&{\small 1.14}&{\small 3.53}$_{-0.50}^{+0.68}$&{\small 1.28}\\[0.3em]
	{\small UGC 1551}&{\small 2}&{\small 1.3}&{\small 0.75}&{\small 1.89}$_{-0.83}^{+1.02}$&{\small 1.32}$_{-0.31}^{+0.18}$&{\small 0.77}&{\small 1.49}$_{-0.49}^{+1.45}$&{\small 2.39}$_{-0.90}^{+2.45}$&{\small 0.81}&{\small 2.40}$_{-0.19}^{+0.21}$&{\small 0.87}\\[0.3em]
	{\small UGC 4325}&{\small 2}&{\small 4.1}&{\small 2.37}&{\small 1.74}$_{-0.35}$&{\small 2.94}\tablenotemark{a}&{\small1.74}&{\small 2.05}$_{-0.30}$&{\small 2.52}\tablenotemark{a}&{\small 1.08}&{\small 9.52}\tablenotemark{a}&{\small3.46}\\[0.3em]
	{\small DDO 64}&{\small 2}&{\small 3.3}&{\small 1.91}&{\small 9.51}$_{-7.89}$&{\small 7.44}$_{-5.52}$&{\small 5.49}&{\small 8.83}$_{-6.98}$&{\small 7.44}$_{-5.77}$&{\small 6.20}&{\small 8.76}\tablenotemark{a}&{\small 3.18}\\[0.3em]
	{\small F583-1}&{\small 2}&{\small 2.5}&{\small 1.44}&{\small 2.16}$_{-0.53}^{+0.69}$&{\small 2.55}$_{-0.19}^{+0.16}$&{\small 1.46}&{\small 1.87}$_{-0.55}^{+0.92}$&{\small 3.57}$_{-0.77}^{+1.68}$&{\small 1.51}&{\small 4.49}$_{-0.15}^{+ 0.16}$&{\small 1.63}
	\enddata
	\tablenotetext{a}{Lower limits.}
	\tablecomments{These parameters represent the best fitting density profiles required to match the central core density out to the edge of the rotation profile data of \citet{kuziodenaray2008}.  All radii are in kpc.  The scale sizes for the four different profiles, the traditional definition of the core size, vary widely despite having the same inner central density and matching the velocity data.  The new definition of core size is much more consistent among profiles.  Error bars are determined from least squares minimization of the model velocity profile residuals (see Appendix).  Since the fitting algorithm uses \rs~and the power law index as parameters, it is difficult to put error bars on \rc~(which is derived directly from the logarthimic slope).  However, the errors on \rc~should be similar to those on \rs.}
	\label{tab:kdnm}
\end{deluxetable*}
\section{The Core Formation Energy Budget}\label{sec:cfeb}
\begin{figure}
	\centering
	\plotone{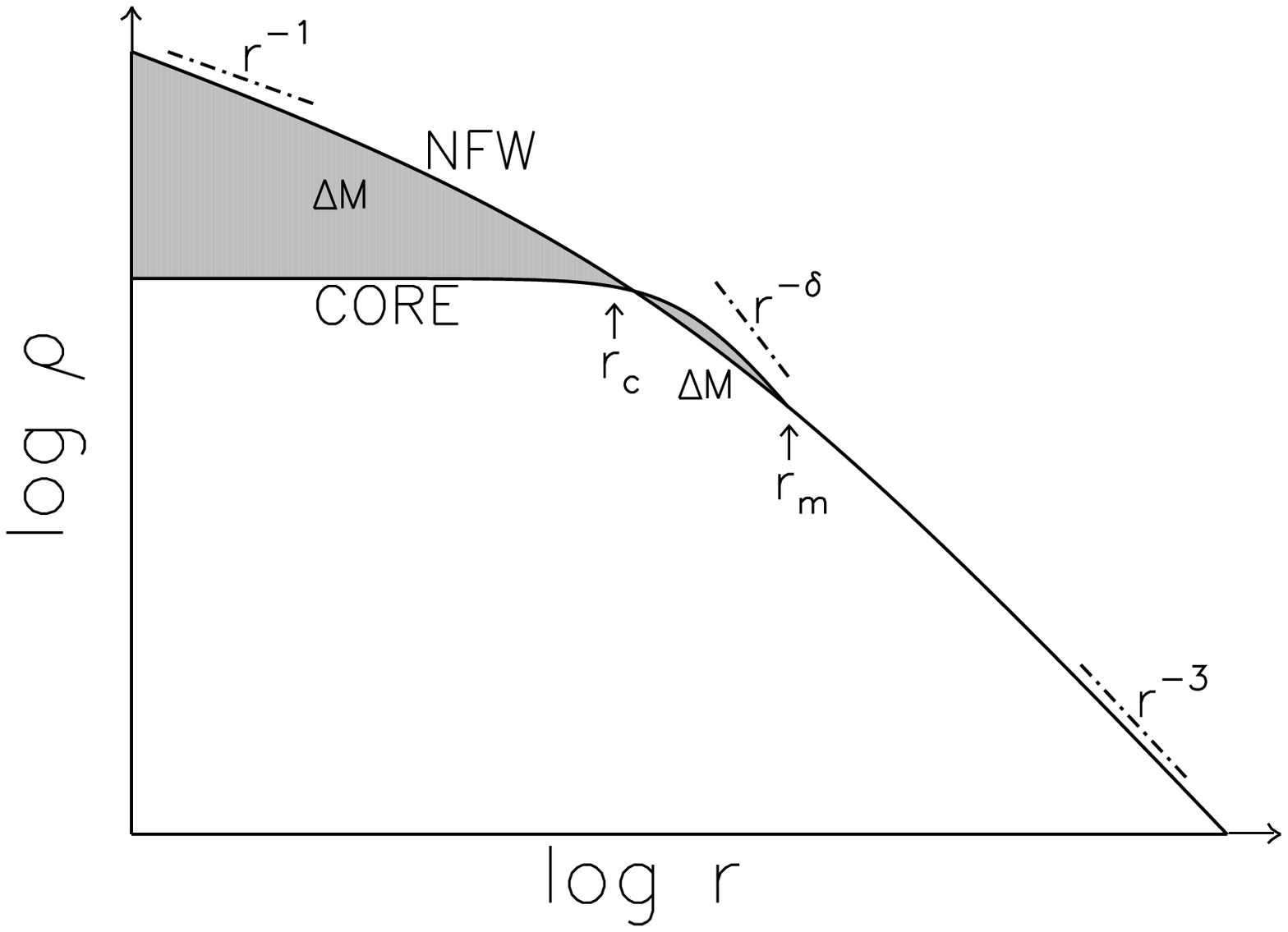}
	\caption[illustration]{A pictorial representation of our procedure to match both density and mass of the core and cusp density profiles at \rmatch.  The leftmost shaded region shows the amount of mass that must be redistributed from the cusp to larger radii (up to \rmatch) in the cored profile. Not having to redistribute mass to larger radii near the virial radius explains why our predictions for the energy cost of core formation are significantly below previous estimates.  It is purely coincidental that in this illustration the core radius, \rc, is close to the radius at which the two density profiles first cross.}
	\label{fig:ill}
\end{figure}

In this section we estimate the energy required to build a core in an initially cuspy density profile.
We assume that each cuspy halo begins with an `NFW' profile \citep[][]{navarro1995, navarro1996, navarro1997}.
If we assume the halos are virialized, we can estimate the energy as the difference between the work required to form the cusp and the work required to form the core, as was done in \citet{penarrubia2012}:
\begin{equation}\label{eq:dE}
	\Delta E=\frac{W_{\rm{c}}-W_{\rm{h}}}{2},
\end{equation}
where \Wh~and \Wc~are the cusp and core potential energies, respectively.
\DE~$>0$ implies the cusped dark matter halo had to gain energy to redistribute mass to form a core.
If the simplifying assumption is made that both halos are spherically symmetric, we can write the potential energy integral as \citep{binney2008}:
\begin{equation}\label{eq:W}
	W=-4\pi G\int\limits_{0}^{\infty}{r\rho(r)M(r)\,\rm{d}r}.
\end{equation}
For a density profile that goes as $\rho\propto r^{-\beta}$ at large $r$, the integral converges for $\beta>5/2$; all of the cored profiles considered (if $\delta > 5/2$ in Eqn. \ref{eq:rhocore}) and the NFW profile lead to well defined values for $W$.\\
\indent Taking the upper limit in Equation \eqref{eq:W} to infinity leads to a very large estimate for the energy (quite apart from the ambiguity of embedding the halo in the cosmic background) and is not appropriate given that we are considering \emph{redistributing} mass within the central part of an existing halo.
\citet{penarrubia2012} chose the halo virial radius \rh~as the upper limit for the potential energy integral, and normalised the mass of their cored density profile,
\begin{equation}\label{eq:rhoppwk}
	\rho(r)=\frac{\rho_{\rm{o}}r_{s}^{3}}{(r_{c}+r)(r_{s}+r)^{2}},
\end{equation}
to that of their initial NFW halo at \rh.
However, simply normalising both halos to have the same mass at \rh~does not remove the requirement that Equation \eqref{eq:W} be carried to infinity. 
In order for the potential energy integral to be valid with a finite upper limit \rmatch, the density \emph{and} mass profiles for both halos must match at that radius.
The choice of \citet{penarrubia2012} to match the mass at \rh~and to truncate the integral for $W$ at that radius builds in an implicit discontinuity in the density at that radius.
Furthermore, moving the cusp mass to near the virial radius of the halo leads to a core formation energy budget from Equations \eqref{eq:W} and \eqref{eq:dE} which will be significantly overestimated compared with the more restricted redistribution considered here (\rmatch $\ll$ \rh).\\
\indent The motivation for constraining the mass and density profiles of the pre- and post-core halos outside \rmatch~to be the same is simply that we expect that strong coupling of star formation feedback to dark matter is only effective in, and only modifies, the central parts of the halo.
The outer parts of the halo will remain largely untouched and reflect the collisionless formation process.
This picture is supported by simulations \citep[e.g.][]{mashchenko2006, mashchenko2008, governato2010, pontzen2012, teyssier2013, madau2014}.
Previous simulations \citep{mashchenko2008, madau2014} suggest that for a $\sim$\thinspace10$^{10}$\thinspace\msol~halo \rmatch~is about 2--3 times the core radius.\\
\indent Our method proceeds as follows:
\begin{itemize}
	\item Since we do not have any robust predictions of what the core profile should be, we adopt Equation \eqref{eq:rhocore} as the density profile when computing \Wc. (We have verified that using a different density profile, such as Equation \eqref{eq:rhoburk} or Equation \eqref{eq:rhofix}, does not change our results.)
	\item We use the relations given by \citet{maccio2007} and \citet{bryan1998} to set the NFW profile parameters at redshift zero as a function of virial halo mass, \Mh~\citep[see][]{penarrubia2012}.  We then have $\rho(r_{\rm{m}})$ and $M(r_{\rm{m}})$ as functions of $M_{\rm h}$.
	\item We find the set of parameters \rhoo, $\delta$, and \rc~for the cored density profile that \emph{simultaneously} satisfy $\rho(r_{\rm{m}})$ and $M(r_{\rm{m}})$ from the previous step.
	\item We set $r_{\rm{m}}=3\,r_{\rm{c}}$, where $r_{\rm c}$ is given by our new definition. Figure \ref{fig:ill} illustrates the profile and mass matching.
	\item We compute the potential energy difference in Eqn~\eqref{eq:dE}.
\end{itemize}
The preceding steps give \DE~in terms of \rc\ and \Mh.
\section{Results}\label{sec:results}
\begin{figure*}
	\centering
	\plotone{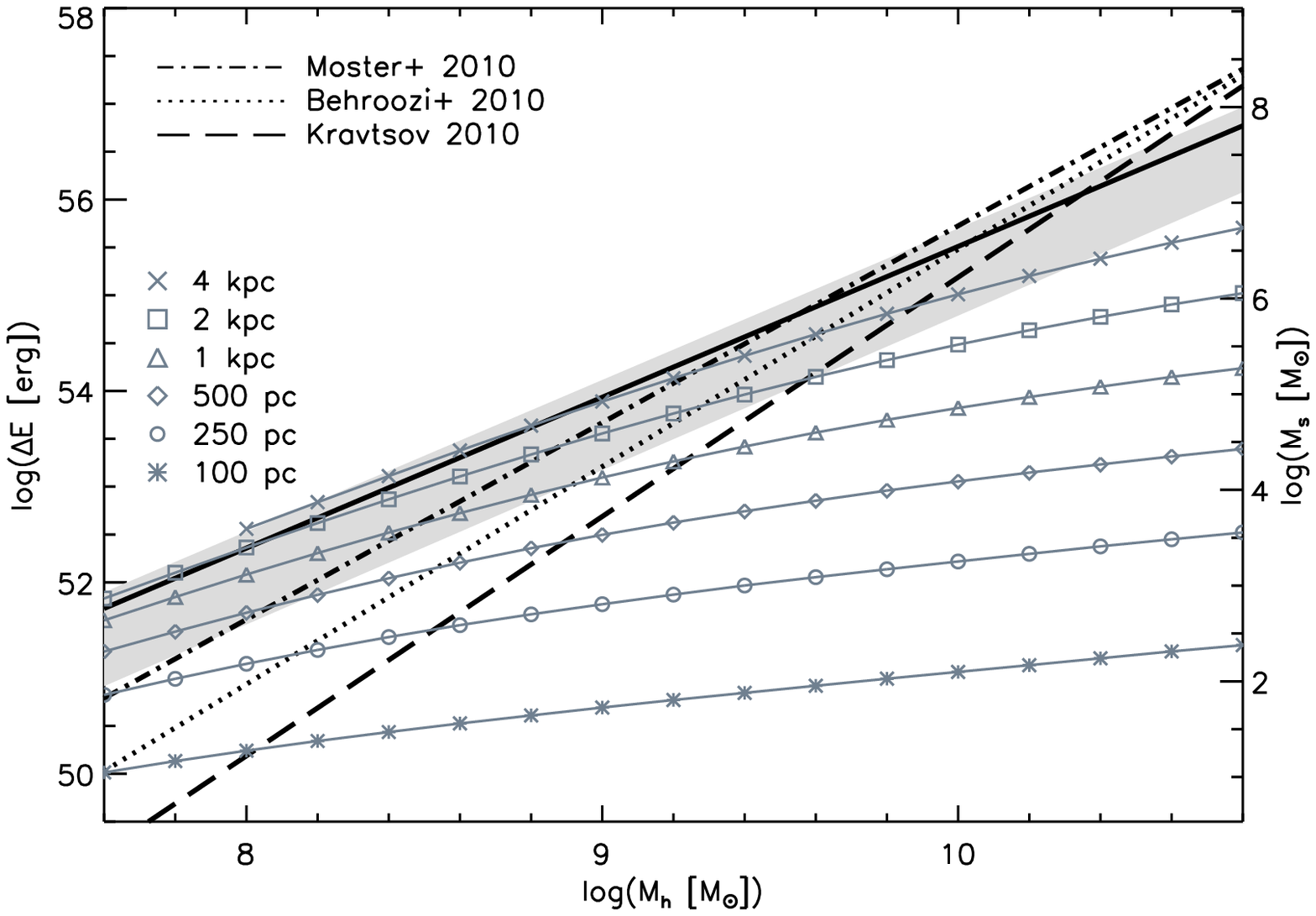}
	\caption[halocomp]{The estimate of the amount of energy \DE~required to convert a dark matter cusp into a core. The solid lines with symbols show the energy required to form a core of a given fixed size using the pseudo-isothermal density profile (see text). The solid black line shows how \DE~scales with \Mh~if \rmatch~is allowed to scale as a fixed fraction of \rh.  The grey area corresponds to the energy estimate of \citet{penarrubia2012}.  The dotted, dashed, and dot-dashed lines correspond to the \Ms-\Mh~relations of \citet{behroozi2010}, \citet{kravtsov2010}, and \citet{moster2010} respectively, assuming a \citet{kroupa2001} IMF.  The right axis shows the stellar mass corresponding to \DE~assuming 100\% efficiency.}
	\label{fig:halocomp}
\end{figure*}

\indent In Figure \ref{fig:halocomp} we show our calculation for the amount of energy required to turn a dark matter cusp into a core as a function of halo mass.
Since we may write approximately:
\begin{equation}\label{eq:Wc}
        W_{\rm{c}}\sim\frac{GM(<r_{\rm{m}})^{2}}{r_{\rm{m}}},
\end{equation}
for the cored profile, we see, as expected, that the parameter that most influences the energy required to convert a cusp into a core is $r_{\rm{m}}\propto r_{\rm{c}}$.
A larger core moves more mass to larger radii and thus requires more energy to redistribute it from the cusp.
On the other hand, if we allow \rmatch~to scale as a fixed fraction of \rh~(we chose \rmatch/\rh=1/6), we find that \DE~scales as \Mh$^{1.6}$.
This scaling is shown by the solid black line in Figure \ref{fig:halocomp} and lies within the range of solutions calculated by \citet{penarrubia2012} (the gray shaded region in the figure).
A more realistic estimate for \rmatch, as a small multiple of \rc~(here 3), leads to a greatly reduced energy requirement and a much shallower dependence on halo mass.
This is shown by the grey lines in Figure \ref{fig:halocomp} for several core sizes.
Our calculations suggest the energy scaling with halo mass is much weaker when \rc~(and hence \rmatch) is fixed.
In the $10^{9}$--$10^{10}$\thinspace\msol~halo mass range, \DE~scales as \Mh$^{0.7}$ for a 1\thinspace kpc core and flattens to \Mh$^{0.3}$ for a 100\thinspace pc core.
This flattening is related to how the mass interior to \rmatch~depends on the halo concentration.
In the functional form we adopted the concentration is inversely proportional to \Mh~\citep{maccio2007}.
Varying the ratio \rmatch/\rc~creates a modest vertical shift in the \DE~curves.\\
\indent In Figure \ref{fig:halocomp} we also plot the stellar mass to halo mass (\Ms-\Mh) relations from \citet{kravtsov2010} (dashed), \citet{behroozi2010} (dotted), and \citet{moster2010} (dash-dot) at $z=0$.
We have converted \Ms~to total SNe energy using a \citet{kroupa2001} IMF, assuming that every star above 8\thinspace\msol~contributes $10^{51}$\thinspace ergs to the energy budget.
Our results show that \emph{there is no conflict between the amount of energy available to form a dark matter core and the $\Lambda$CDM framework}, regardless of halo mass.
The main question is how large the dark matter core is likely to be.\\
\indent In reality, not all of the SNe energy will couple to the cusp dark matter through the induced perturbations to the gravitational potential.
This inefficiency can be represented by scaling the \Ms-\Mh~relations shown in Figure \ref{fig:halocomp} by some coupling efficiency factor, $\epsilon$, \citep[for example, the 40\% value used by][]{penarrubia2012}.
Only for $\epsilon\lesssim0.01\%$ would there be no possibility of creating a core larger than 250\thinspace pc in the 10$^{9}$--10$^{10}$\thinspace\msol~halo range.
The implication is that most, if not all, dwarf galaxies should host dark matter cores if their central star formation rate was high enough to generate 10$^{3}$--10$^{4}$\thinspace\msol~in stars.\\
\subsection{Efficiency of Core Formation}\label{sec:disc}
\begin{figure}
	\centering
	\epsscale{1.2}
	\plotone{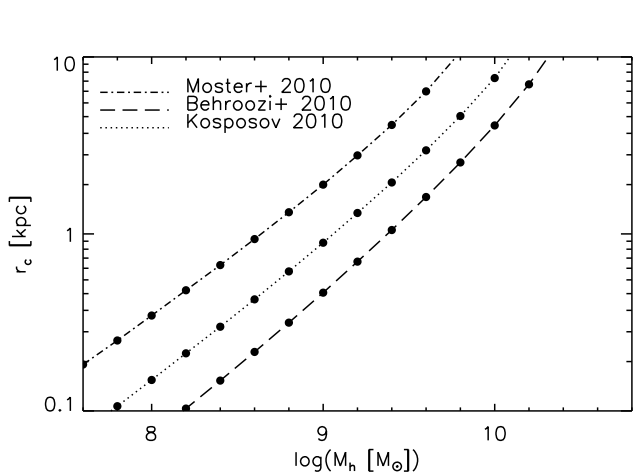}
	\caption[halocore]{The predicted core size as a function of halo mass using the three \Ms-\Mh~relations discussed in this paper.  It is extremely difficult to compare these predictions to observed dark matter cores in local group dwarf galaxies given the uncertainties in the relations, and the difficulty in determining \Mh~for individual galaxies.}
	\label{fig:halocore}
\end{figure}

\begin{figure}
	\centering
	\epsscale{1.2}
	\plotone{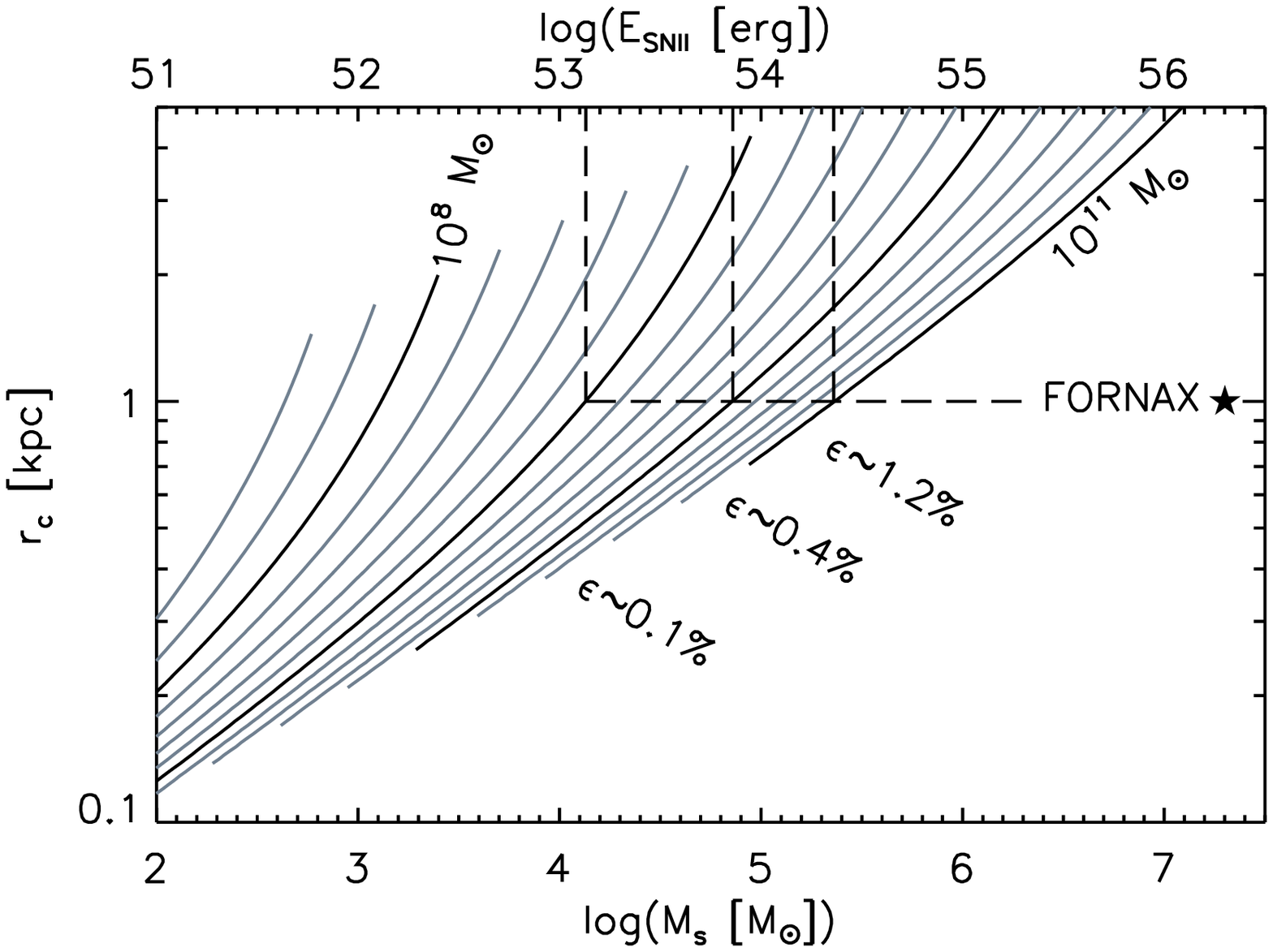}
	\caption[stellarcore]{The predicted dark matter core \emph{as a function of stellar mass}.  The grey lines show the energy curve in \rc-\Ms~space for a given \Mh.  The top axis shows our assumed conversion from \Ms\ to SNe energy.  For clarity, we have highlighted the curves for halos of $10^{8}$\thinspace\msol, $10^{9}$\thinspace\msol, $10^{10}$\thinspace\msol, and $10^{11}$\thinspace\msol.  The star shows the stellar mass ($2\times10^{7}$\thinspace\msol) of Fornax from \citet{mcconnachie2012} with the dark matter core size estimated by \citet{amorisco2013}.  The efficiency $\epsilon$ to which the SNe feedback couples to the creation of a dark matter core can be estimated by sliding the Fornax data point to the left.}
	\label{fig:stellarcore}
\end{figure}

In the previous section we showed that stellar feedback, of which we have only considered SNe, can provide enough energy to the dense star forming gas in the centres of dwarf galaxies to transform the dark matter cusps predicted by numerical experiments to the flattened dark matter cores inferred from observations \citep[e.g.][]{mashchenko2008, pontzen2012, maxwell2012}.
This is contrary to previous work \citep[e.g.][]{penarrubia2012} and results from our adoption of a more consistent mass normalisation criterion.
This was motivated in part by the results of hydrodynamic simulations and by the observation that the most efficient injection of star formation feedback energy will occur at the deepest point in the galactic potential, which implies a physical limit to the radius to which cusp dark matter can be redistributed.
In our calculations, we have parameterized this limit as the radius \rmatch.
Our results suggest significant dark matter cores can be formed with less than 1\% of the total SNe energy available over the star forming history of a typical dwarf galaxy.\\
\indent So far we have been concerned with whether a dark matter core can form in an initially cuspy halo due to star formation feedback.
Our new calculations allow us to ask a more interesting question: what is the typical core size for a given halo mass?
In Figure \ref{fig:halocore} we show the predicted core size as a function of halo mass using the three \Ms-\Mh~relations shown in Figure \ref{fig:halocomp} and 100\% of the associated SNe energy.
Although Figure \ref{fig:halocore} reinforces our main conclusion that dark matter cores can form from cusps in even the smallest dwarf galaxy halos, it also emphasizes the uncertainty in trying to make detailed predictions.
The variation in the low mass slope of the \Ms-\Mh~relation between the \citet{moster2010}, \citet{behroozi2010}, and \citet{kravtsov2010} results corresponds directly to a factor of 2--4 variation in predicted \rc~as a function of halo mass.
Furthermore, these relations only give the average stellar mass as a function of halo mass.\\
\indent Figure \ref{fig:halocore} is further complicated because we have not yet considered the efficiency with which the energy from stellar feedback is transferred to dark matter particle orbits.
The coupling efficiency will likely vary in time with changing conditions in the centre of the halo, and will depend on the star formation rate, gas accretion rate, gas heating rate, and the dark matter particle orbits; the value of $\epsilon$ used in \citet{penarrubia2012} is merely a time-averaged value over the dwarf galaxy star formation and gas accretion histories.
Finally, our results show that the energy required to form the core, \DE, has only a weak dependence on \Mh~for a given \rc.\\
\indent If, indeed, our ansatz that matter redistribution happens only within \rmatch~is correct, then the total halo mass \Mh~has only a modest influence on the final core sizes in dwarf galaxies.
Many observations of dwarf galaxies in the local universe can only measure dynamical masses out to a few kpc \citep[e.g.][]{dalcanton2009, oh2011a, kuziodenaray2006, kuziodenaray2008, mcconnachie2012} and so cannot accurately determine the halo virial mass.
This leads us to Figure \ref{fig:stellarcore} where we have plotted the range of core sizes as a function of stellar mass required to produce a core of that size, \emph{without} adopting a specific coupling efficiency in dark matter core formation.
Each curve represents the track in \rc-\Ms~space for each halo mass, and is determined by the relations between \Mh, \rh, and \rs~(or concentration parameter).
The endpoints of each curve represent the upper and lower limits that we imposed when computing the energy requirements from Equation \eqref{eq:W}.\\
\indent In this figure, the efficiency $\epsilon$ is equivalent to sliding a given dwarf galaxy along the \Ms~axis.
To illustrate this, we show the local group dwarf Fornax, whose stellar mass was taken from \citet{mcconnachie2012} and whose dark matter core size was measured by \citet{amorisco2013}.
Fornax lies significantly below the predicted core size for its stellar mass at 100\% efficiency, but is consistent with our core predictions if only a few per cent of the total SNe budget contributed to the formation of the dark matter core in the host halo.
Fornax's host halo could range anywhere from 10$^{9}$--10$^{11}$\thinspace\msol~while $\epsilon$ could range anywhere from 0.1--1\%.\\
\section{Discussion and Conclusions}
In this paper we presented a new estimate for the energy requirement for galaxies to form dark matter cores in their host halos.
We assume former cusp material remains within a characteristic radius, \rmatch, at which both the density and mass profiles of the cored dark matter halo and the original cusped halo match.
This significantly reduces the amount of energy required compared to the estimate of \citet{penarrubia2012} without having to require, for example, that cores form only at high redshift \citep[e.g.][]{amorisco2014}.
Our results suggest that dark matter cores are present in most, if not all, dark matter halos that experienced clustered star formation within their centres \citep{maxwell2012}.
In the case of the Local Group dwarf galaxy Fornax, if it lives in a $10^{10}$\thinspace\msol~halo then less than 1\% of the energy released by all supernovae over its star formation history would be required to build a 1 kpc core.\\
\indent Our new calculations alleviate another potential problem raised by \citeauthor{penarrubia2012}: the accretion of additional high phase-space density collisionless particles (cuspy satellites) was shown by \citet{dehnen2005} to tend to preserve the steepest density cusp in the merger.
However, if all large halos merging in the collisionless mass assembly process are already cored, the dark matter density profile of the merger remnant should be cored as well.
While the smallest accretors, common at early times, may be cuspy and rebuild a central cusp, subsequent star formation feedback should be enough to destroy the dark matter cusp in a short time\footnote{The work of \citet{mashchenko2008}, for example, showed that a core can form in $\sim100$\thinspace Myr.}.\\
\indent With the apparent tension between $\Lambda$CDM and the presence of cores in dwarf galaxies relieved, we can now focus on making firm predictions of the sizes of dark matter cores as a function of stellar mass (Figure \ref{fig:stellarcore}).
For example, simulations of dark matter core formation in dwarf galaxies can test the efficiency and energy output of star formation prescriptions.
The resulting density profiles and other properties can be directly compared to observations of local dwarfs \citep[e.g.][]{kuziodenaray2008, dalcanton2009}.
It is also important to note that we have neglected all other forms of star formation feedback aside from supernovae \citep[e.g.][]{hopkins2011, hopkins2013, shen2010, agertz2013, keller2014}.
However, these extra sources of energy can simply be approximated by tuning $\epsilon$, the efficiency with which the energy is transferred to the dark matter, using the energy from supernovae as a baseline.\\
\indent The ability for a given halo to form a core will depend on the coupling between the depth of the central halo potential and the star formation rate which limits the amount of available feedback energy.
The bursty star formation at the centres of dwarf galaxies should be able to drive $\epsilon$ high enough to form a sizeable core \citep{maxwell2012} at high redshift \citep{mashchenko2008, amorisco2014, madau2014}.
However, low mass halos may not be able to form cores if their galactic potential is shallow enough that each burst of star formation results in completely efficient gas expulsion \citep[e.g.][]{navarro1996b, read2005, shen2014, madau2014}.\\[10pt]
\indent The authors wish to thank NSERC for support and R. Kuzio De Naray for providing the observed rotation profile data.

\appendix
\section{Velocity Profile Fits}
\begin{figure}
	\centering
	\epsscale{2.1}
	\plottwo{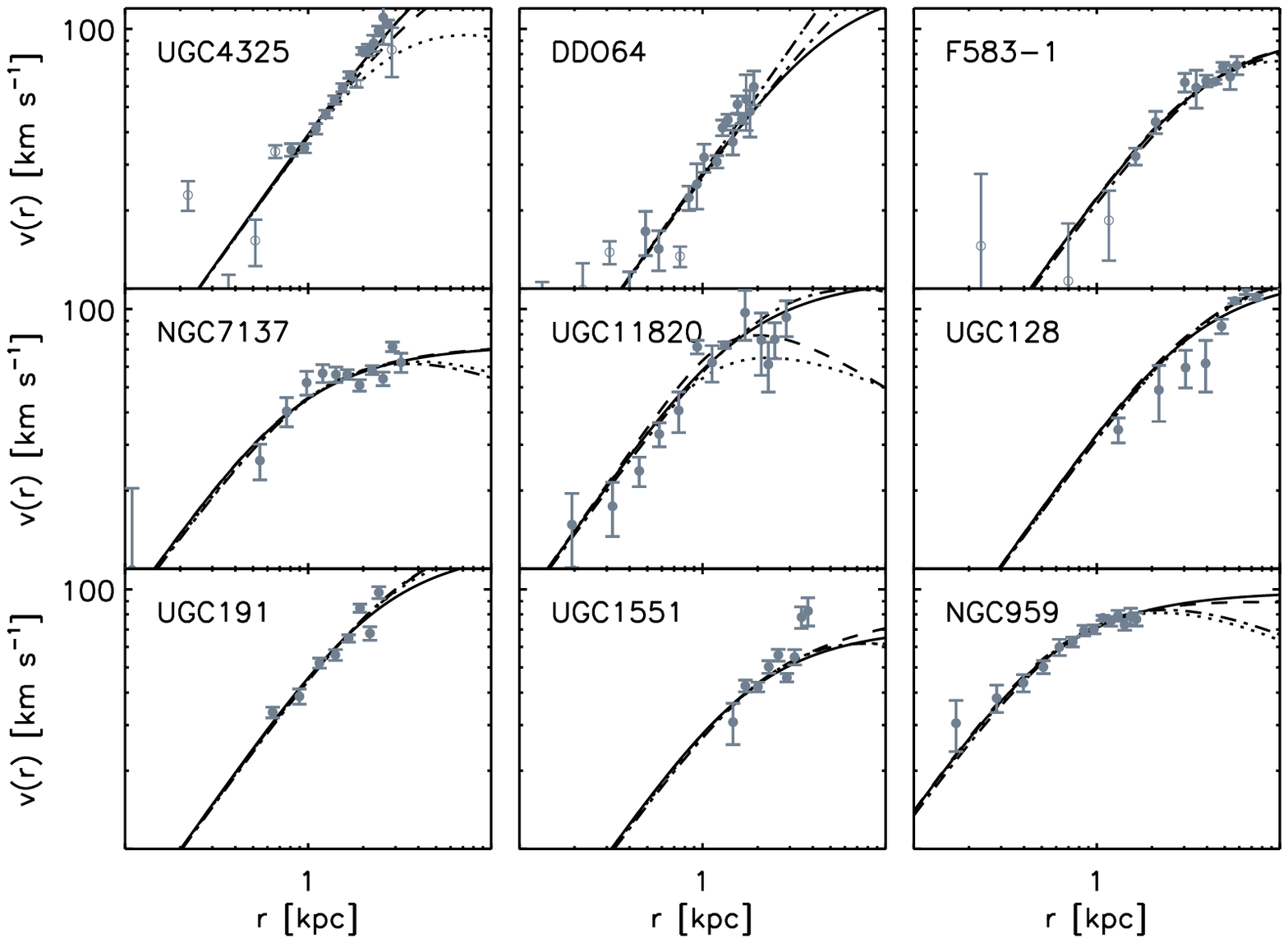}{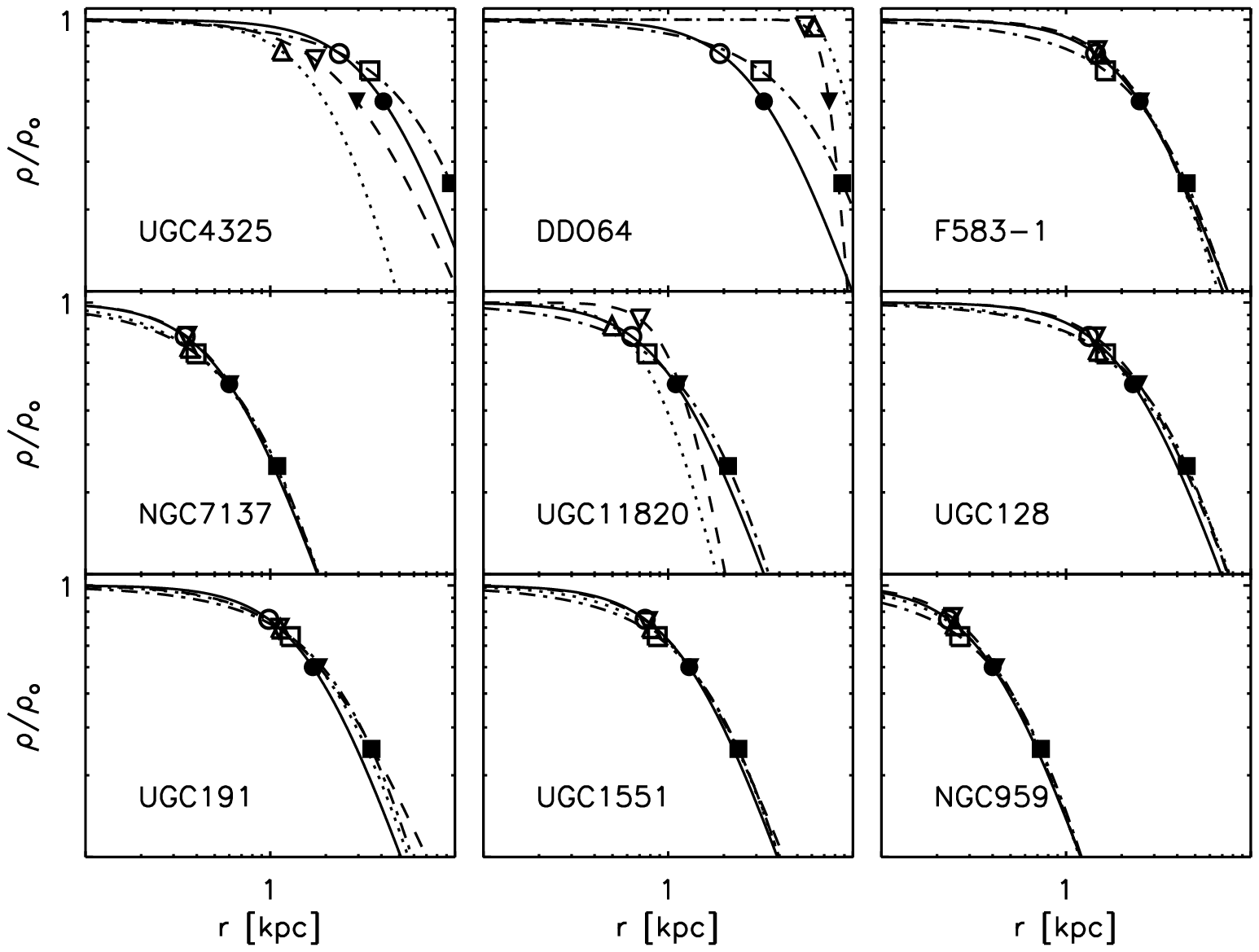}
	\label{fig:kdnm}
	\caption{The velocity and density profiles for all of the \citet{kuziodenaray2008} LSBs.  Line styles and symbols are the same as in Figure \ref{fig:kdn}.}
\end{figure}

Figure \ref{fig:kdnm} repeats the analysis shown in Figure \ref{fig:kdn} for all nine Low Surface Brightness galaxies observed by \citet{kuziodenaray2008}.
\section{Error Estimation for Best Fit Values}
\begin{figure}
   \centering
   \epsscale{1.0}
   \plotone{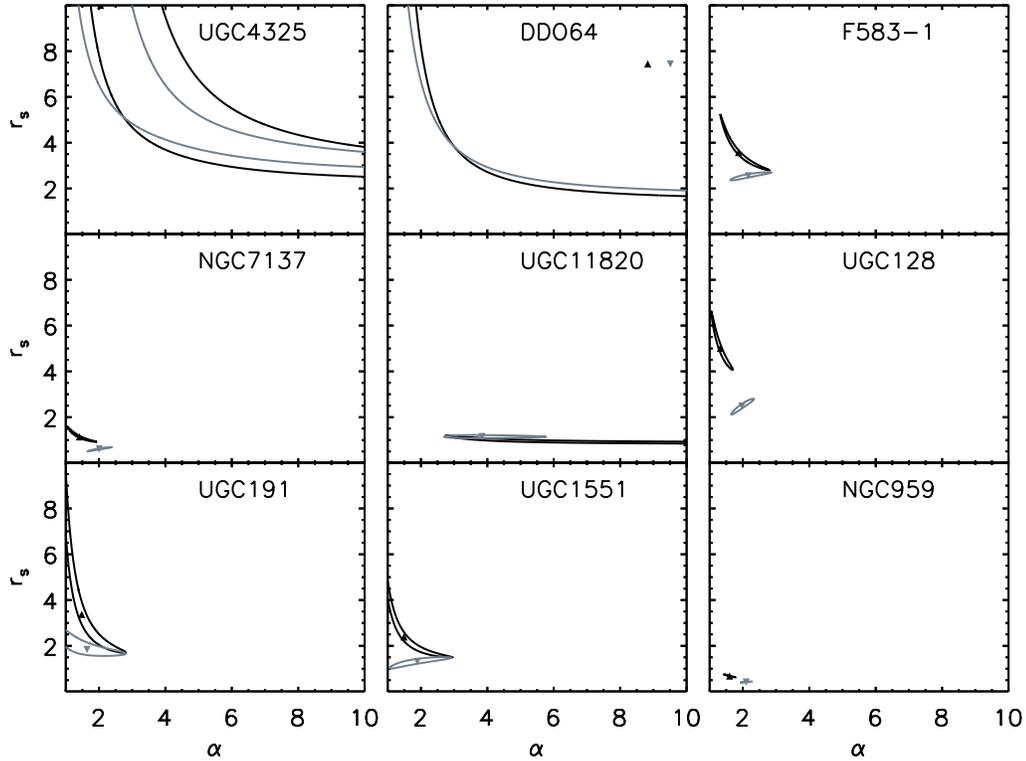}
	\label{fig:kdnc}
   \caption{The uncertainty contours for all of the \citet{kuziodenaray2008} LSBs, with black and gray indicating the pseudo-isothermal and fixed slope profiles, respectively.}
\end{figure}

We determine the best fit by finding the minimum in the square of the velocity residual, $\mathcal{R}^{2}$:
\begin{equation}\label{eq:R}
	\mathcal{R}^{2}=\sum_{i=1}^{N}{\left(\frac{R_{i}}{\Delta_{i}}\right)^{2}},
\end{equation}
where $R_{i}$ is the velocity residual at each measured radius and $\Delta_{i}$ is the measured velocity error from \citet{kuziodenaray2008}.
If the density models used in the text were linear in the parameters \rs, $\alpha$, and $\delta$, the uncertainties could be derived by using the $\chi^{2}$ probability distribution to determine confidence levels.
Since this is not the case, the probability distribution of $\mathcal{R}^{2}$ cannot simply be approximated to that of $\chi^{2}$.
Instead, we compute an uncertainty contour given by:
\begin{equation}
	\sigma^{2}=\frac{\text{min}(\mathcal{R}^{2})}{N-M},
\end{equation}
where $M$ is the number of free parameters --- two for Equations \eqref{eq:rhofix} and \eqref{eq:rhocore} and one for Equation \eqref{eq:rhoburk} --- since the central density is fixed.
In Figure \ref{fig:kdnc}, these contours are shown by the solid black and gray lines for Equations \eqref{eq:rhofix} and \eqref{eq:rhocore}, respectively, for the sample of nine galaxies from \citet{kuziodenaray2008}.
Velocity data points that showed large variations from the expected velocity profile were ignored when computing the error contours; these are shown as open circles in Figure \ref{fig:kdnm}.
The location of the minimum $\mathcal{R}^{2}$ for each  are given by the filled triangles.
For most of these galaxies, minimizing the residuals gives relatively tight constraints on the scale radius and power index for the density profiles.
However, UGC 4325 and DDO 64 do not have closed contours, which is unsurprising given that constraining \rs~requires the velocity profile to begin to turn over --- traits that are not exhibited in the velocity data for these galaxies.
Instead, we can only define acceptable lower limits for the density profile parameters.

\begin{thebibliography}{}
\expandafter\ifx\csname natexlab\endcsname\relax\def\natexlab#1{#1}\fi
\bibitem[{{Adams} {et~al.}(2014){Adams}, {Simon}, {Fabricius}, {van den Bosch}, {Barentine}, {Bender}, {Gebhardt}, {Hill}, {Murphy}, {Swaters}, {Thomas}, \& {van de Ven}}]{adams2014}
{Adams}, J.~J., {Simon}, J.~D., {Fabricius}, M.~H., {et~al.} 2014, \apj, 789, 63
\bibitem[{{Agertz} {et~al.}(2013){Agertz}, {Kravtsov}, {Leitner}, \& {Gnedin}}]{agertz2013}
{Agertz}, O., {Kravtsov}, A.~V., {Leitner}, S.~N., \& {Gnedin}, N.~Y. 2013, \apj, 770, 25
\bibitem[{{Amorisco} {et~al.}(2013){Amorisco}, {Agnello}, \& {Evans}}]{amorisco2013}
{Amorisco}, N.~C., {Agnello}, A., \& {Evans}, N.~W. 2013, \mnras, 429, L89
\bibitem[{{Amorisco} {et~al.}(2014){Amorisco}, {Zavala}, \& {de Boer}}]{amorisco2014}
{Amorisco}, N.~C., {Zavala}, J., \& {de Boer}, T.~J.~L. 2014, \apjl, 782, L39
\bibitem[{{Behroozi} {et~al.}(2010){Behroozi}, {Conroy}, \& {Wechsler}}]{behroozi2010}
{Behroozi}, P.~S., {Conroy}, C., \& {Wechsler}, R.~H. 2010, \apj, 717, 379
\bibitem[{{Binney} \& {Tremaine}(2008)}]{binney2008}
{Binney}, J., \& {Tremaine}, S. 2008, {Galactic Dynamics: Second Edition} (Princeton University Press)
\bibitem[{{Bosma}(1981)}]{bosma1981}
{Bosma}, A. 1981, \aj, 86, 1825
\bibitem[{{Bryan} \& {Norman}(1998)}]{bryan1998}
{Bryan}, G.~L., \& {Norman}, M.~L. 1998, \apj, 495, 80
\bibitem[{{Bullock} {et~al.}(2001){Bullock}, {Kolatt}, {Sigad}, {Somerville}, {Kravtsov}, {Klypin}, {Primack}, \& {Dekel}}]{bullock2001}
{Bullock}, J.~S., {Kolatt}, T.~S., {Sigad}, Y., {et~al.} 2001, \mnras, 321, 559
\bibitem[{{Burkert}(1995)}]{burkert1995}
{Burkert}, A. 1995, \apjl, 447, L25
\bibitem[{{C{\^o}t{\'e}} {et~al.}(2000){C{\^o}t{\'e}}, {Carignan}, \& {Freeman}}]{cote2000}
{C{\^o}t{\'e}}, S., {Carignan}, C., \& {Freeman}, K.~C. 2000, \aj, 120, 3027
\bibitem[{{Dalcanton} {et~al.}(2009){Dalcanton}, {Williams}, {Seth}, {Dolphin}, {Holtzman}, {Rosema}, {Skillman}, {Cole}, {Girardi}, {Gogarten},  {Karachentsev}, {Olsen}, {Weisz}, {Christensen}, {Freeman}, {Gilbert}, {Gallart}, {Harris}, {Hodge}, {de Jong}, {Karachentseva}, {Mateo}, {Stetson}, {Tavarez}, {Zaritsky}, {Governato}, \& {Quinn}}]{dalcanton2009}
{Dalcanton}, J.~J., {Williams}, B.~F., {Seth}, A.~C., {et~al.} 2009, \apjs, 183, 67
\bibitem[{{Dehnen}(2005)}]{dehnen2005}
{Dehnen}, W. 2005, \mnras, 360, 892
\bibitem[{{Dubinski} {et~al.}(2009){Dubinski}, {Berentzen}, \& {Shlosman}}]{dubinski2009}
{Dubinski}, J., {Berentzen}, I., \& {Shlosman}, I. 2009, \apj, 697, 293
\bibitem[{{Dubinski} \& {Carlberg}(1991)}]{dubinski1991}
{Dubinski}, J., \& {Carlberg}, R.~G. 1991, \apj, 378, 496
\bibitem[{{El-Zant} {et~al.}(2001){El-Zant}, {Shlosman}, \& {Hoffman}}]{elzant2001}
{El-Zant}, A., {Shlosman}, I., \& {Hoffman}, Y. 2001, \apj, 560, 636
\bibitem[{{Flores} \& {Primack}(1994)}]{flores1994}
{Flores}, R.~A., \& {Primack}, J.~R. 1994, \apjl, 427, L1
\bibitem[{{Gilmore} {et~al.}(2007){Gilmore}, {Wilkinson}, {Wyse}, {Kleyna}, {Koch}, {Evans}, \& {Grebel}}]{gilmore2007}
{Gilmore}, G., {Wilkinson}, M.~I., {Wyse}, R.~F.~G., {et~al.} 2007, \apj, 663, 948
\bibitem[{{Governato} {et~al.}(2010){Governato}, {Brook}, {Mayer}, {Brooks}, {Rhee}, {Wadsley}, {Jonsson}, {Willman}, {Stinson}, {Quinn}, \& {Madau}}]{governato2010}
{Governato}, F., {Brook}, C., {Mayer}, L., {et~al.} 2010, \nat, 463, 203
\bibitem[{{Governato} {et~al.}(2012){Governato}, {Zolotov}, {Pontzen}, {Christensen}, {Oh}, {Brooks}, {Quinn}, {Shen}, \& {Wadsley}}]{governato2012}
{Governato}, F., {Zolotov}, A., {Pontzen}, A., {et~al.} 2012, \mnras, 422, 1231
\bibitem[{{Hogan} \& {Dalcanton}(2000)}]{hogan2000}
{Hogan}, C.~J., \& {Dalcanton}, J.~J. 2000, \prd, 62, 063511
\bibitem[{{Hopkins} {et~al.}(2011){Hopkins}, {Quataert}, \& {Murray}}]{hopkins2011}
{Hopkins}, P.~F., {Quataert}, E., \& {Murray}, N. 2011, \mnras, 417, 950
\bibitem[{{Jardel} \& {Sellwood}(2009)}]{jardel2009}
{Jardel}, J.~R., \& {Sellwood}, J.~A. 2009, \apj, 691, 1300
\bibitem[{{Keller} {et~al.}(2014){Keller}, {Wadsley}, {Benincasa}, \& {Couchman}}]{keller2014}
{Keller}, B.~W., {Wadsley}, J., {Benincasa}, S.~M., \& {Couchman}, H.~M.~P. 2014, \mnras, 442, 3013
\bibitem[{{Klypin} {et~al.}(2001){Klypin}, {Kravtsov}, {Bullock}, \& {Primack}}]{klypin2001}
{Klypin}, A., {Kravtsov}, A.~V., {Bullock}, J.~S., \& {Primack}, J.~R. 2001, \apj, 554, 903
\bibitem[{{Kravtsov}(2010)}]{kravtsov2010}
{Kravtsov}, A. 2010, AdAst, 2010
\bibitem[{{Kroupa}(2001)}]{kroupa2001}
{Kroupa}, P. 2001, \mnras, 322, 231
\bibitem[{{Kuzio de Naray} {et~al.}(2008){Kuzio de Naray}, {McGaugh}, \& {de Blok}}]{kuziodenaray2008}
{Kuzio de Naray}, R., {McGaugh}, S.~S., \& {de Blok}, W.~J.~G. 2008, \apj, 676, 920
\bibitem[{{Kuzio de Naray} {et~al.}(2006){Kuzio de Naray}, {McGaugh}, {de Blok}, \& {Bosma}}]{kuziodenaray2006}
{Kuzio de Naray}, R., {McGaugh}, S.~S., {de Blok}, W.~J.~G., \& {Bosma}, A. 2006, \apjs, 165, 461
\bibitem[{{Leitherer} {et~al.}(1999){Leitherer}, {Schaerer}, {Goldader}, {Delgado}, {Robert}, {Kune}, {de Mello}, {Devost}, \& {Heckman}}]{leitherer1999}
{Leitherer}, C., {Schaerer}, D., {Goldader}, J.~D., {et~al.} 1999, \apjs, 123, 3
\bibitem[{{Ma} \& {Boylan-Kolchin}(2004)}]{ma2004}
{Ma}, C.-P., \& {Boylan-Kolchin}, M. 2004, \prl, 93, 021301
\bibitem[{{Macci{\`o}} {et~al.}(2007){Macci{\`o}}, {Dutton}, {van den Bosch}, {Moore}, {Potter}, \& {Stadel}}]{maccio2007}
{Macci{\`o}}, A.~V., {Dutton}, A.~A., {van den Bosch}, F.~C., {et~al.} 2007, \mnras, 378, 55
\bibitem[{{Madau} {et~al.}(2014){Madau}, {Shen}, \& {Governato}}]{madau2014}
{Madau}, P., {Shen}, S., \& {Governato}, F. 2014, \apjl, 789, L17
\bibitem[{{Mashchenko} {et~al.}(2006){Mashchenko}, {Couchman}, \& {Wadsley}}]{mashchenko2006}
{Mashchenko}, S., {Couchman}, H.~M.~P., \& {Wadsley}, J. 2006, \nat, 442, 539
\bibitem[{{Mashchenko} {et~al.}(2008){Mashchenko}, {Wadsley}, \& {Couchman}}]{mashchenko2008}
{Mashchenko}, S., {Wadsley}, J., \& {Couchman}, H.~M.~P. 2008, Science, 319, 174
\bibitem[{{Maxwell} {et~al.}(2012){Maxwell}, {Wadsley}, {Couchman}, \& {Mashchenko}}]{maxwell2012}
{Maxwell}, A.~J., {Wadsley}, J., {Couchman}, H.~M.~P., \& {Mashchenko}, S. 2012, \apjl, 755, L35
\bibitem[{{Maxwell} {et~al.}(2014){Maxwell}, {Wadsley}, {Couchman}, \& {Sills}}]{maxwell2014}
{Maxwell}, A.~J., {Wadsley}, J., {Couchman}, H.~M.~P., \& {Sills}, A. 2014, \mnras, 439, 2043
\bibitem[{{McConnachie}(2012)}]{mcconnachie2012}
{McConnachie}, A.~W. 2012, \aj, 144, 4
\bibitem[{{Moore}(1994)}]{moore1994}
{Moore}, B. 1994, \nat, 370, 629
\bibitem[{{Moster} {et~al.}(2010){Moster}, {Somerville}, {Maulbetsch}, {van den Bosch}, {Macci{\`o}}, {Naab}, \& {Oser}}]{moster2010}
{Moster}, B.~P., {Somerville}, R.~S., {Maulbetsch}, C., {et~al.} 2010, \apj, 710, 903
\bibitem[{{Navarro} {et~al.}(1996{\natexlab{a}}){Navarro}, {Eke}, \& {Frenk}}]{navarro1996b}
{Navarro}, J.~F., {Eke}, V.~R., \& {Frenk}, C.~S. 1996{\natexlab{a}}, \mnras, 283, L72
\bibitem[{{Navarro} {et~al.}(1995){Navarro}, {Frenk}, \& {White}}]{navarro1995}
{Navarro}, J.~F., {Frenk}, C.~S., \& {White}, S.~D.~M. 1995, \mnras, 275, 720
\bibitem[{{Navarro} {et~al.}(1996{\natexlab{b}}){Navarro}, {Frenk}, \& {White}}]{navarro1996}
---. 1996{\natexlab{b}}, \apj, 462, 563
\bibitem[{{Navarro} {et~al.}(1997){Navarro}, {Frenk}, \& {White}}]{navarro1997}
---. 1997, \apj, 490, 493
\bibitem[{{Oh} {et~al.}(2011){Oh}, {de Blok}, {Brinks}, {Walter}, \& {Kennicutt}}]{oh2011a}
{Oh}, S.-H., {de Blok}, W.~J.~G., {Brinks}, E., {Walter}, F., \& {Kennicutt}, Jr., R.~C. 2011, \aj, 141, 193
\bibitem[{{Pe{\~n}arrubia} {et~al.}(2012){Pe{\~n}arrubia}, {Pontzen}, {Walker}, \& {Koposov}}]{penarrubia2012}
{Pe{\~n}arrubia}, J., {Pontzen}, A., {Walker}, M.~G., \& {Koposov}, S.~E. 2012, \apjl, 759, L42
\bibitem[{{Pontzen} \& {Governato}(2012)}]{pontzen2012}
{Pontzen}, A., \& {Governato}, F. 2012, \mnras, 421, 3464
\bibitem[{{Pontzen} \& {Governato}(2014)}]{pontzen2014}
---. 2014, \nat, 506, 171
\bibitem[{{Read} \& {Gilmore}(2005)}]{read2005}
{Read}, J.~I., \& {Gilmore}, G. 2005, \mnras, 356, 107
\bibitem[{{Rubin} {et~al.}(1980){Rubin}, {Ford}, \& {.~Thonnard}}]{rubin1980}
{Rubin}, V.~C., {Ford}, W.~K.~J., \& {.~Thonnard}, N. 1980, \apj, 238, 471
\bibitem[{{Rubin} {et~al.}(1978){Rubin}, {Thonnard}, \& {Ford}}]{rubin1978}
{Rubin}, V.~C., {Thonnard}, N., \& {Ford}, Jr., W.~K. 1978, \apjl, 225, L107
\bibitem[{{Sellwood}(2003)}]{sellwood2003}
{Sellwood}, J.~A. 2003, \apj, 587, 638
\bibitem[{{Shen} {et~al.}(2014){Shen}, {Madau}, {Conroy}, {Governato}, \& {Mayer}}]{shen2014}
{Shen}, S., {Madau}, P., {Conroy}, C., {Governato}, F., \& {Mayer}, L. 2014, \apj, 792, 99
\bibitem[{{Shen} {et~al.}(2010){Shen}, {Wadsley}, \& {Stinson}}]{shen2010}
{Shen}, S., {Wadsley}, J., \& {Stinson}, G. 2010, \mnras, 407, 1581
\bibitem[{{Spergel} \& {Steinhardt}(2000)}]{spergel2000}
{Spergel}, D.~N., \& {Steinhardt}, P.~J. 2000, \prl, 84, 3760
\bibitem[{{Stadel} {et~al.}(2009){Stadel}, {Potter}, {Moore}, {Diemand}, {Madau}, {Zemp}, {Kuhlen}, \& {Quilis}}]{stadel2009}
{Stadel}, J., {Potter}, D., {Moore}, B., {et~al.} 2009, \mnras, 398, L21
\bibitem[{{Tasitsiomi}(2003)}]{tasitsiomi2003}
{Tasitsiomi}, A. 2003, \ijmpd, 12, 1157
\bibitem[{{Teyssier} {et~al.}(2013){Teyssier}, {Pontzen}, {Dubois}, \& {Read}}]{teyssier2013}
{Teyssier}, R., {Pontzen}, A., {Dubois}, Y., \& {Read}, J.~I. 2013, \mnras, 429, 3068
\bibitem[{{Walker} \& {Pe{\~n}arrubia}(2011)}]{walker2011}
{Walker}, M.~G., \& {Pe{\~n}arrubia}, J. 2011, \apj, 742, 20
\bibitem[{{Weinberg} \& {Katz}(2002)}]{weinberg2002}
{Weinberg}, M.~D., \& {Katz}, N. 2002, \apj, 580, 627
\bibitem[{{Widrow}(2000)}]{widrow2000}
{Widrow}, L.~M. 2000, \apjs, 131, 39
\bibitem[{{Zolotov} {et~al.}(2012){Zolotov}, {Brooks}, {Willman}, {Governato}, {Pontzen}, {Christensen}, {Dekel}, {Quinn}, {Shen}, \& {Wadsley}}]{zolotov2012}
{Zolotov}, A., {Brooks}, A.~M., {Willman}, B., {et~al.} 2012, \apj, 761, 71
\end{thebibliography}
\end{document}